\documentclass[pdflatex,sn-mathphys-num]{sn-jnl}

\usepackage{amssymb}
\usepackage{amsmath}
\usepackage{bm}
\usepackage{natbib}
\usepackage[export]{adjustbox}

\begin{document}

\title{MeltpoolINR: Predicting temperature field, melt pool geometry, and their rate of change in laser powder bed fusion}

\author*[1]{\fnm{Manav} \sur{Manav}}\email{mmanav@ethz.ch}\equalcont{These authors contributed equally to this work.}

\author[2]{\fnm{Nathanaël} \sur{Perraudin}}\equalcont{These authors contributed equally to this work.}

\author[3]{\fnm{Yunong} \sur{Lin}}\equalcont{These authors contributed equally to this work.}

\author[3]{\fnm{Mohamadreza} \sur{Afrasiabi}}

\author[2]{\fnm{Fernando} \sur{Perez-Cruz}}

\author*[3]{\fnm{Markus} \sur{Bambach}}\email{mbambach@ethz.ch}

\author*[1]{\fnm{Laura} \sur{De Lorenzis}}\email{ldelorenzis@ethz.ch}

\affil[1]{\orgdiv{Computational Mechanics Group}, \orgname{ETH Zürich}, \orgaddress{\city{Zürich}, \country{Switzerland}}}

\affil[2]{\orgdiv{Swiss Data Science Center}, \orgname{ETH Zürich}, \orgaddress{\city{Zürich}, \country{Switzerland}}}

\affil[3]{\orgdiv{Advanced Manufacturing Lab}, \orgname{ETH Zürich}, \orgaddress{\city{Zürich}, \country{Switzerland}}}

\abstract{
We present a data-driven, differentiable neural network model designed to learn the temperature field, its gradient, and the cooling rate, while implicitly representing the melt pool boundary as a level set in laser powder bed fusion. The physics-guided model combines fully connected feed-forward neural networks with Fourier feature encoding of the spatial coordinates and laser position. Notably, our differentiable model allows for the computation of temperature derivatives with respect to position, time, and process parameters using autodifferentiation. Moreover, the implicit neural representation of the melt pool boundary as a level set enables the inference of the solidification rate and the rate of change in melt pool geometry relative to process parameters. The model is trained to learn the top view of the temperature field and its spatiotemporal derivatives during a single-track laser powder bed fusion process, as a function of three process parameters, using data from high-fidelity thermo-fluid simulations. The model accuracy is evaluated and compared to a state-of-the-art convolutional neural network model, demonstrating strong generalization ability and close agreement with high-fidelity data.
}

\keywords{Laser powder bed fusion, Melt pool, Deep learning, Implicit neural representation}

\newcommand{\vect}[1]{\bm{#1}}
\newcommand{\fref}[1]{Figure~\ref{#1}}
\newcommand{\sref}[1]{Section~\ref{#1}}
\newcommand{\tref}[1]{Table~\ref{#1}}

\maketitle

\section{Introduction}
\label{introduction}
Laser powder bed fusion (LPBF) is a prominent metal additive manufacturing process for metal parts with complex geometries. It offers remarkable flexibility in fabricating customized parts with significantly reduced lead times compared to conventional manufacturing processes. For these reasons, it finds increasing application in biomedical, aerospace, automotive, and other high-tech industries~\cite{attaran2017rise,yadroitsev2021fundamentals,afrasiabi2023modelling}. 
In LPBF, part fabrication progresses through the sequential spreading of powder layers and their selective melting via a laser, leading to the formation of a melt pool, and subsequent solidification. Thus, the printing process involves complex multiscale and multiphysics phenomena, including laser beam-powder interaction and melt pool formation, melt pool dynamics, and fast cooling and solidification cycles~\cite{king2015laser}. 

The process is controlled by adjusting a number of parameters, including laser power, scanning speed, and preheating temperature, as well as the scanning strategy~\cite{oliveira2020processing}. 
However, the inherent complexity of the involved physical mechanisms obscures the direct correlation between the process parameters and the quality of a fabricated part; thus, ensuring precise control over part quality remains challenging, rendering the printed parts susceptible to defects such as keyhole porosity, lack of fusion, and surface defects~\cite{snow2020invited,mostafaei2022defects}.

This has motivated researchers to deepen their understanding of the underlying physics of the printing process~\cite{king2015laser,soundararajan2021review} and to develop reliable models and numerical methods to simulate, optimize, and control LPBF with computational tools~\cite{jardin2023optimizing,pham2023framework,mccann2021situ,wang2020model}. In this direction, the first analytical model of the temperature field due to a moving point source, akin to the LPBF fabrication process, was developed by Rosenthal~\cite{rosenthal1946theory}. Semi-analytical models of temperature field due to a moving heat source by incorporating idealized boundary effects have also been developed~\cite{yang2018semi}. These models are computationally efficient, but their accuracy is limited by strong simplifying assumptions such as neglecting the temperature dependence of material properties, the latent heat of fusion and the heat loss due to radiation and convection, and other simplifications regarding boundary conditions. Significant progress has also been made in detailed simulation of the LPBF process~\cite{markl2016multiscale,moges2019review,soundararajan2021review} using complex mathematical models discretized with numerical methods such as finite elements (FE)~\cite{luo2018survey,dunbar2016experimental}, finite volumes~\cite{acharya2017prediction,yu2016role}, smoothed particle hydrodynamics (SPH)~\cite{afrasiabi2021multi,furstenau2020generating,luthi2023adaptive}, Lattice Boltzmann~\cite{zakirov2020predictive}, and others. These approaches have simulated specific parts of the printing process and provided crucial insights into the physics of the process and explanations for the experimental observations. However, high-fidelity simulations are computationally very expensive, prohibiting their wider application in process optimization and control.

In recent years, machine learning models have seen tremendous growth in various scientific applications. They have been employed in learning from data as well as in the direct numerical solution of partial differential equations~\cite{karniadakis2021physics}. These models can be used for quick inference, enabling the possibility of optimization and control. This has led to a surge in research focused on employing machine learning for LPBF process modeling~\cite{wang2022process,mozaffar2022mechanistic}. In recognition of the importance of the temperature field during the printing process on the part quality, considerable effort has been directed toward machine learning models that can predict the temperature field in and around the melt pool. The models presented in these works fall into two main categories: physics-informed and data-driven. A physics-informed approach that models heat transfer and melt pool fluid dynamics is presented in~\cite{zhu2021machine}. This model is not parametric, i.e. it solves the governing equations only for a prescribed set of process parameters like in a direct numerical simulation. A nonparametric PINNs approach solving only the heat conduction equation for the LPBF process is proposed in~\cite{liao2023hybrid}. The authors also solve an inverse problem using PINNs to identify material properties from the partially observed temperature data. A parametric solution of the heat conduction equation for the LPBF process is obtained using PINNs in~\cite{hosseini2023single}. The proposed PINNs approaches have either modeled the physics adequately or obtained a parametric solution to only the heat conduction equation, but not both. Among the data-driven approaches, \cite{chen2022data} proposes a method to learn the temperature field around the melt pool by learning a parametrization of isotherms and reconstruction of the temperature field through interpolation using data from thermal-fluid flow simulations. In~\cite{ren2020thermal}, the temperature field obtained from the FE solution of only the heat transfer equation for a variety of scanning paths is learned using a combination of a recurrent neural network and a fully connected feed-forward network. Similarly obtained temperature field data for multilayer printing are learned using a fully connected feed-forward network in~\cite{pham2023fast}. To accurately learn the temperature field in the melt pool from the high-fidelity data obtained from a computational fluid dynamics simulation, a model based on a convolutional neural network (CNN) is employed in~\cite{hemmasian2023surrogate}. In~\cite{sideris2023gpyro}, an uncertainty-aware temperature model is proposed by combining physics-informed and parametric regressors utilizing heat transfer equation and experimental data.

In the LPBF process, the solidification of the melt pool typically entails high cooling rates and large thermal gradients which depend on the process parameters~\cite{bertoli2017situ}. Notably, the cooling rates and thermal gradients near the melt pool boundary define the solidification rate and have significant influence on the microstructure~\cite{lienert2011fundamentals,wei2015evolution,debroy2018additive} and on residual stresses~\cite{kruth2012assessing,li2018residual,fang2020review}, and consequently on the mechanical properties and geometric accuracy of the fabricated part~\cite{levkulich2019effect}. The construction of a solidification map~\cite{kou2003welding,blecher2014solidification} also relies on the knowledge of solidification rate and thermal gradient near the melt pool boundary. Furthermore, for a gradient-based optimization of the process parameters, the temperature field should be represented as a sufficiently smooth function of the spatiotemporal coordinates and of the process parameters, such that its derivatives can be computed accurately. Hence, a model that accurately predicts the temperature field, melt pool boundary, thermal gradient and cooling rate at the melt pool boundary, and their change due to a change in the process parameters is desired. However, none of the works in the literature has focused on the development of such a model. Notably, (semi)-analytical models yield such relations~\cite{rosenthal1946theory,yang2018semi}, however, their accuracy is limited. As mentioned earlier, direct numerical solutions are not parametric and can provide accurate fields for only a prescribed set of process parameters, besides being computationally expensive. Parametric PINNs-based surrogate models~\cite{liao2023hybrid,hosseini2023single} can predict temperature field, thermal gradient, and cooling rate accurately and fast; however, they have not been employed for the prediction of melt pool evolution and its dependence on the process parameters. Furthermore, the reported approaches model physics only partially i.e. they solve only the thermal conduction equation. Among the data-driven machine learning approaches~\cite{chen2022data,ren2020thermal,pham2023fast,hemmasian2023surrogate,sideris2023gpyro}, focus has been on learning the temperature field and not the thermal gradient and cooling rate. Also, they have not been applied to predict the melt pool evolution and its dependence on the process parameters.

In this paper, we propose an efficient data-driven approach by developing a differentiable model based on fully connected feed-forward neural networks, commonly known as multilayer perceptrons (MLPs), for learning the temperature field,  thermal gradient, and cooling rate, inspired by the implicit neural representation of 3D objects~\cite{mildenhall2021nerf}. We call our model MeltpoolINR which stands for ``melt pool implicit neural representation''. Unlike CNNs, MeltpoolINR learns a resolution-independent continuous representation of the temperature field with advantages in the computation of the derivatives of the field and of the melt pool boundary. The inputs to the model are spatial coordinates, laser position, and the process parameters. We also employ Fourier feature encoding of coordinates and laser position which are well known to enable MLPs to learn sharp features in the data~\cite{tancik2020fourier} such as in the temperature field and its derivatives near the melt pool boundary. Furthermore, to accurately capture the extremely sharp temperature variation at the front of the melt pool boundary, we include an exponential term in the expression for the temperature inspired by Rosenthal's solution for a moving heat source~\cite{rosenthal1946theory}. Additionally, we employ the level-set method to represent the melt pool boundary implicitly and to model its evolution. We perform Sobolev training~\cite{czarnecki2017sobolev} of our model to learn the temperature field and its derivatives accurately using the high-fidelity physics-based data generated from SPH process simulations. In this work, we demonstrate the effectiveness of our approach in learning a 2D cross-section (top view) of the temperature field and its derivatives in single-track printing. The novel contributions of this work are threefold: (a) development of a differentiable neural network capable of learning the temperature field, thermal gradient and cooling rate, and providing the derivative of the temperature with respect to the process parameters using auto-differentiation, (b) implicit neural representation of the melt pool boundary as a level set to learn its evolution and infer solidification rate, and (c) inference of the rate of change of melt pool features with respect to the process parameters.

The remainder of this paper is organized as follows. The data generation method is briefly outlined in \sref{sph}. The machine learning model and the level-set method are described in \sref{method}. The results are presented and discussed in \sref{result}. The paper ends with conclusions in \sref{conclusions}.


\section{Data generation}
\label{sph}
The dataset for training and validating the machine learning model is obtained by SPH numerical simulations. SPH, introduced in \cite{gingold1977smoothed} and \cite{lucy1977numerical}, is a particle-based numerical method widely used to solve differential equations describing large deformations, fluid flow, and phase transformations. In SPH, the continuum body is discretized into finite particles, whereby each particle interacts with its neighbors according to the governing physical equations. Here, we use a thermo-fluid SPH approach, combined with a ray-tracing laser model introduced in \cite{LIN2023124378}, to simulate the single-track laser scanning process. As follows, we summarize the mathematical model and SPH discretization. For full details see \cite{LIN2023124378, luthi2023adaptive}.

\begin{figure}[h]
	\centering
	\includegraphics[scale=0.45]{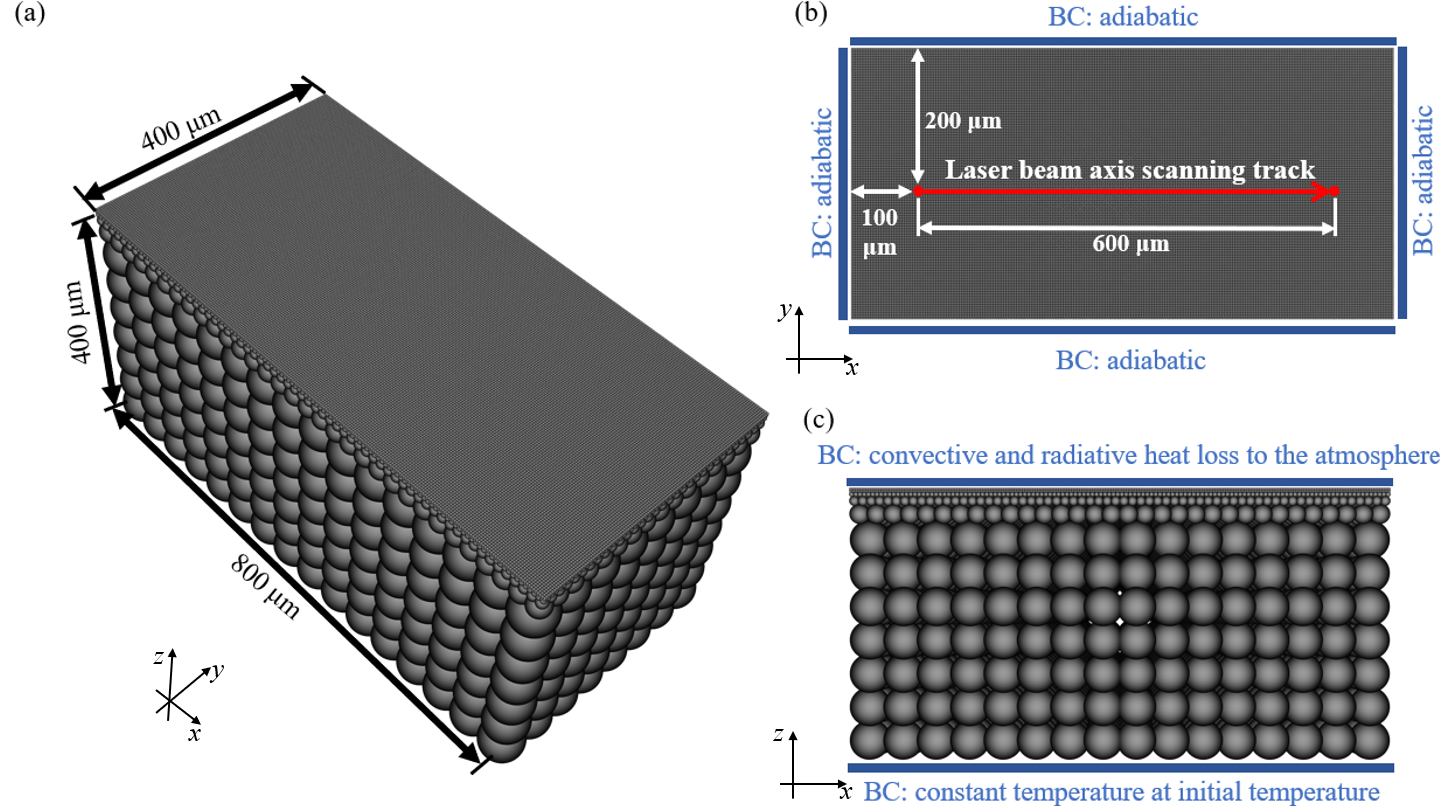}
	\caption{SPH simulation setup: (a) geometry of the substrate; (b) boundary conditions on the lateral faces and laser beam axis scanning track; (c) boundary conditions on the top and bottom faces.}
	\label{fig:sim_model}
\end{figure}

We simulate a laser scanning process over a single track, i.e. the elementary process into which an entire LPBF process can be decomposed. In total 210 simulations are performed for different values of the process parameters, i.e. laser power ($P$), laser scanning speed ($v$), and initial substrate temperature ($T_{pre}$), using all combinations of the values in \tref{tab:values}. Here, the laser power and the laser scanning speed can be set in the LPBF machine for the scanning of individual tracks, whereas the initial substrate temperature represents the substrate heat accumulation by previous scanning. In the present study, we do not account for the randomly distributed particle geometry in the powder bed, in order to limit the computational cost. After the simulation, a post-processing step extracts the desired fields at the top surface from the SPH numerical results. The data for each case have the same dimension of [100, 128, 64] wherein 100 is the number of time steps and [128, 64] are the numbers of grid points in $x$ and $y$ directions (\fref{fig:sim_model}), respectively.
\begin{table}[h]
	\caption{Values of the process parameters in the dataset.}
	\centering
	\begin{tabular}{l |l }
		\hline
		Laser power ($P$) {[}W{]}      & 80, 120, 160, 200, 240, 280         \\ \hline
		Laser scanning speed ($v$) {[}m/s{]}   & 0.75, 1, 1.25, 1.5, 1.75, 2.25, 2.5 \\ \hline
		Initial substrate temperature ($T_{pre}$) {[}K{]} & 300, 340, 380, 420, 460             \\ \hline
	\end{tabular}
	\label{tab:values}
\end{table}

In every simulated single-track process, a Gaussian laser beam with spot diameter 80 $\mu$m pointing vertically downwards scans the top surface of a Ti6Al4V substrate. The laser power and scanning speed do not change during each process. The substrate is simplified as a cuboid of dimensions $800\times400\times400$ $\mu$m, as shown in \fref{fig:sim_model}a. The laser axis moves along the midline of the top surface of the substrate in the $y$ direction, as depicted in \fref{fig:sim_model}b. The scanning path is 600 $\mu$m long and its starting point is positioned with an offset of 100 $\mu$m to the left edge of the top surface. At the beginning of the process, the substrate has a uniform initial temperature. The material properties of the Ti6Al4V alloy for the simulations are listed in \tref{tab:Ti64 properties}. The thermal conductivity and isobaric specific heat capacity are temperature-dependent and estimated using their values in the solid and liquid phases.

\begin{table}[ht!]
	\caption{Material properties of Ti6Al4V alloy and atmospheric conditions of the simulated process~\cite{LIN2023124378, Solidus_liquidus}.}
	\centering
	\begin{tabular}{l|c |c |c }
		\hline
		Property / Atmospheric condition & Symbol & Value & Unit  \\
		\hline
		Reference density  & $\rho_0$ & 4420 & kgm$^{-3}$ \\       
		Solid thermal conductivity constant & $k_{S}$ & 33.4 & Wm$^{-1}$K$^{-1}$ \\
		Liquid thermal conductivity constant & $k_{L}$ & 19  & Wm$^{-1}$K$^{-1}$ \\
		\begin{tabular}[c]{@{}l@{}}Solid isobaric specific heat capacity \\ constant\end{tabular} & $c_{p_{S}}$ & 830 & Jkg$^{-1}$K$^{-1}$ \\
		\begin{tabular}[c]{@{}l@{}}Liquid isobaric specific heat capacity \\ constant\end{tabular}& $c_{p_{L}}$ & 546  & Jkg$^{-1}$K$^{-1}$\\
		Dynamic viscosity  & $\mu$ & 0.005 & kgm$^{-1}$s$^{-1}$ \\
		Surface tension  & $\sigma_T$ & 1.493 & Nm$^{-1}$ \\    
		Surface tension gradient coefficient & $\sigma_T'$ & $1.9\times 10^{-4}$ & Nm$^{-1}$K$^{-1}$\\     
		Solidus temperature & $T_{S}$ & 1878 & K \\       
		Liquidus temperature & $T_{L}$ & 1933 & K \\   
		Reference temperature of evaporation  & $T_v$ & 3133 & K\\       
		Specific enthalpy of melting  & $H_{m}$ & $2.9 \times 10^{5}$ & Jkg$^{-1}$ \\ 
		Specific enthalpy of evaporation  & $H_v$ & $9.83 \times 10^{6}$ & Jkg$^{-1}$ \\   
		Atmospheric pressure  & $P_{atm}$ & $10^{5}$ & Pa\\     
		Atmospheric temperature  & $T_{atm}$ & 300 & K \\       
		Convective heat transfer coefficient  & $h_c$ & 19 & Wm$^{-2}$K$^{-1}$  \\ 
		Emissivity  & $\epsilon$ & 0.23 & $-$ \\       
		Molar mass  & $M_m$ & 0.048 & kgmol$^{-1}$\\    
		\hline
	\end{tabular}
	\label{tab:Ti64 properties}
\end{table}

We solve the equations of incompressible fluid flow and heat transfer in the Lagrangian form:
\begin{equation}
	\left\{
	\begin{aligned}
		\frac{d\rho}{dt} & = -\rho\nabla \cdot \mathbf{u} \\
		\rho \frac{d\mathbf{u}}{dt} & = -\nabla p + \mu \nabla^2 \mathbf{u} + \rho \mathbf{g} + \mathbf{b}^s + \mathbf{b}^w + \mathbf{b}^r\\
		\rho c_p \frac{dT}{dt} & =\nabla \cdot (k \nabla T) + \mathbf{T} :\nabla \mathbf{u} + q^{rc} + q^l + (q^{b})\\
	\end{aligned}
	\right.
	\label{data_eq1}
\end{equation}
Here $\rho$ denotes the density, $t$ the time, $\mathbf{u}$ the velocity, $p$ the pressure, $\mu$ the dynamic viscosity, and $\mathbf{g}$ the gravity acceleration. $\mathbf{b}^s$, $\mathbf{b}^w$ and $\mathbf{b}^r$ are the body forces induced respectively by surface tension, wetting effect, and recoil pressure of evaporation. $c_p$ is the isobaric specific heat capacity, $k$ the thermal conductivity, $T$ the temperature, and $\mathbf{T}$ the shear stress tensor. $q^{rc}$ is the heat loss induced by evaporation, $q^{l}$ the laser heat input which is computed by a ray-tracing laser heat source model, and $q^{b}$ is the heat loss induced by boundary conditions which is only applied to the surface.

As depicted in \fref{fig:sim_model}b and \fref{fig:sim_model}c, the four vertical faces of the substrate are subjected to adiabatic boundary conditions, hence $q^{b}=0$. The top face experiences convective and radiative heat loss to the environment, with the atmospheric conditions in \tref{tab:Ti64 properties}, hence
\begin{equation}
	q^{b} = -\frac{h_c(T-T_{atm}) + \epsilon \sigma (T^4 - T_{atm}^4)}{\Delta x}  
\end{equation}
where $h_c$ is the convective heat loss coefficient, $T_{atm}$ the atmospheric temperature, $\epsilon$ the emissivity, $\sigma$ the Stefan–Boltzmann constant, and $\Delta x$ the minimum discretization spacing. The bottom face maintains a constant initial temperature. In the simulation, this boundary condition is achieved by adding two layers of ghost SPH particles below the substrate. During the time integration, the temperature of the ghost SPH particles remains the initial substrate temperature. The ghost particles thus conduct heat from interior SPH particles.

To reduce the computational cost, we deploy an adaptive spatial resolution scheme in SPH which refines and coarsens the particle cloud at selected time integration steps. The minimum discretization spacing of SPH is 6.3 $\mu$m, and the time integration step is 6.3 ns.

\begin{figure}[h]
	\centering
	\includegraphics[scale=0.59]{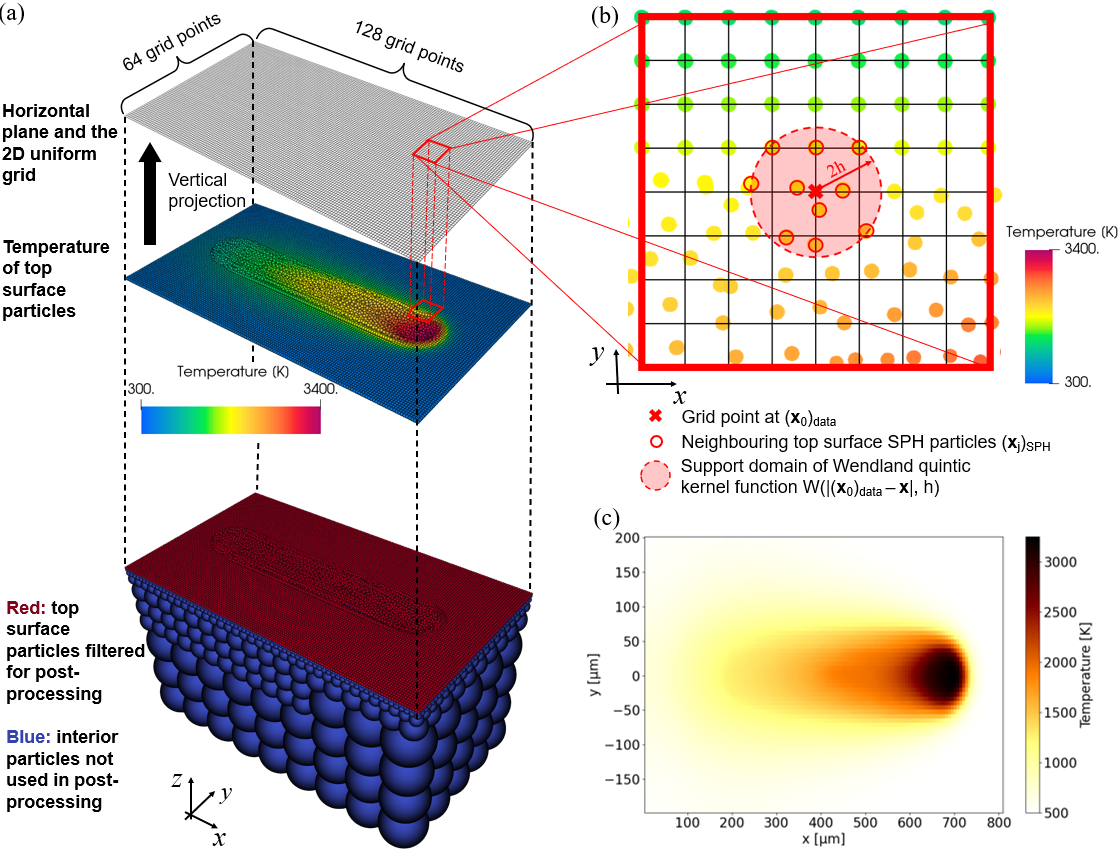}
	\caption{Generation of temperature data for the dataset from SPH particles at one time frame; (a) vertical projection of filtered top surface particles to a uniform grid on a 2D plane; (b) computation of temperature at a grid point $(\mathbf{x}_0)_{data}$ by SPH interpolation; (c) post-processed temperature field data of dimension [128, 64] at one time frame.}
	\label{fig:SPH_postprocess}
\end{figure}
The post-processing of the numerical solution follows three steps. First, from the SPH simulations, the fields at the 100 uniformly spaced time frames covering the whole process are saved to generate the 2D field data. Secondly, as \fref{fig:SPH_postprocess}a depicts, the top surface particles for ray-tracing, which are determined by the method in \cite{LIN2023124378}, are filtered out and projected vertically onto a horizontal plane. A uniform $128\times64$ grid along the $x$ and $y$ directions is created on the horizontal plane. Thirdly, the 2D data consisting of the field variables on the grid points are computed by a 2D SPH interpolation with a renormalization denominator from the surface particles as follows: 
\begin{equation}
	f(\mathbf{x}_0)_{data} = \frac{\sum_{j} f(\mathbf{x}_j)_{SPH} W[|(\mathbf{x}_0)_{data}-(\mathbf{x}_j)_{SPH}|,h)](V_j)_{SPH}}
	{\sum_{j} W[|(\mathbf{x}_0)_{data}-(\mathbf{x}_j)_{SPH}|,h)](V_j)_{SPH}}
	\label{eq:SPH_function_interpolation}
\end{equation}
Here, $f(\mathbf{x}_0)_{data}$ is the value of a field variable at position $(\mathbf{x}_0)_{data}$, $j$ denotes a particle in the neighborhood of $(\mathbf{x}_0)_{data}$ with position $(\mathbf{x}_j)_{SPH}$, and $W[\cdot,\cdot]$ is the kernel function, a weighting function whose value depends on $|(\mathbf{x}_0)_{data}-(\mathbf{x}_j)_{SPH}|$ and on the smoothing length $h$. The neighboring particles of $(\mathbf{x}_0)_{data}$ are the particles in the support domain of the kernel function $W(|(\mathbf{x}_0)_{data}-\mathbf{x}|,h)$. $(V_j)_{SPH}$ is the volume of particle $j$. Here, we use as kernel function the 2D Wendland quintic function presented in \cite{Wendland} with a smoothing length of 4.725 $\mu$m. \fref{fig:SPH_postprocess}b schematically shows the relationship between top surface SPH particles and grid point at $(\mathbf{x}_0)_{data}$ in the SPH interpolation. Finally, at each time frame, field variables including the temperature $T_{data}$, the temperature gradient in $x$-direction $(\partial_x T)_{data}$, the temperature gradient in y direction $(\partial_y T)_{data}$, and the temporal temperature gradient $(\partial_t T)_{data}$ at the 2D uniform grid are generated (\fref{fig:SPH_postprocess}c).

\section{Methodology}
\label{method}
In this section, we describe first our machine learning approach, and then the level set method, useful for the implicit representation of the melt pool.

\subsection{Machine learning approach}
\label{dl}
\begin{figure}[!ht]
	\centering
	\includegraphics[width=0.9\textwidth]{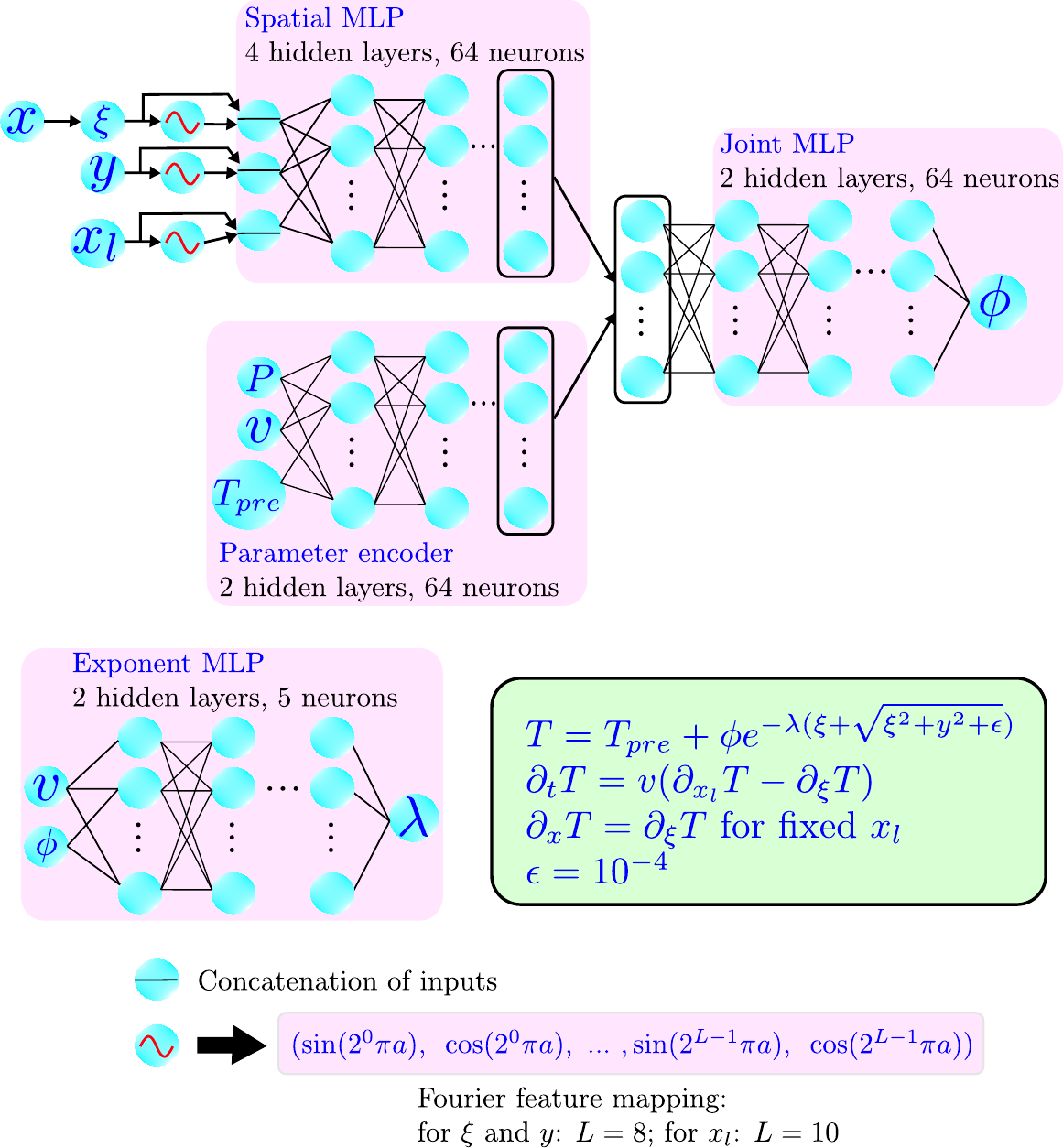}
	\caption{The architecture of MeltpoolINR with the Fourier feature mapping of the spatial coordinates and laser position.}
	\label{fig:MLP_architecture}
\end{figure}

We learn the top view of the temperature field, its gradient, and its time derivative in a single-track LPBF using a physics-guided Fourier feature network~\cite{mildenhall2021nerf,tancik2020fourier,garbin2021fastnerf} based on MLPs, which we denote as MeltpoolINR (Implicit Neural Representation of the melt pool) (\fref{fig:MLP_architecture}). The architecture consists of four MLPs each employing the hyperbolic tangent ($\tanh$) activation function:
\begin{enumerate}
	\item Spatial MLP: The inputs to this MLP are the transformed spatial coordinates of points in the printing domain and the laser position, $x_l$. First, the $x$-coordinate of points is transformed to the coordinate attached to the laser moving in the $x$-direction by setting $\xi=x-x_l$. Then Fourier feature mapping is applied to $\xi$, $y$, and $x_l$ and the input to the network is constructed by appending values of $\xi$, $y$, and $x_l$ to their corresponding Fourier feature maps. The network consists of 4 hidden layers with 64 neurons in each layer.
	\item Parameter encoder: This network takes the process parameters as input and encodes them in a latent space. It consists of 2 hidden layers with 64 neurons in each layer.
	\item Joint MLP: The outputs of the Spatial MLP and the Parameter encoder are concatenated and supplied as input to this network. This network also consists of 2 hidden layers with 64 neurons in each layer.
	\item Exponent MLP: Rosenthal's solution for a moving heat source~\cite{rosenthal1946theory} involves an exponential term in which the exponent includes laser velocity and thermal diffusivity. Here, we learn the exponent using a small MLP which consists of 2 hidden layers with 5 neurons in each layer.
\end{enumerate}

The temperature field and its derivatives vary sharply near the melt pool boundary which is challenging for an MLP to learn. To overcome this, we augment the NN architecture with two features: (a) Fourier feature mapping, and (b) the inclusion of an exponential term in the output based on Rosenthal's solution for a moving heat source~\cite{rosenthal1946theory}. Fourier feature mapping enables MLPs to learn high-frequency features and is extensively employed in computer vision for this purpose~\cite{mildenhall2021nerf,tancik2020fourier,garbin2021fastnerf}. Furthermore, the sharply changing temperature field and its derivatives at the front of the melt pool still pose a challenge. To mitigate this, we include an exponential term in the expression for the temperature (see~\fref{fig:MLP_architecture}) and learn the exponent from the data using an MLP. Note that the small parameter $\epsilon$ in the expression for the temperature smoothens the corner point, which would otherwise appear at $\xi=y=0$. Furthermore, to enforce the physical constraint that the time derivative should be equal to the negative of the x-component of the temperature gradient times the scanning speed in the steady state (see~\ref{frame_change}), we learn the temperature field in the coordinate system attached to the laser by setting $\xi=x-x_l$. The separation of MLPs for the dense spatial coordinates and the sparse parameter set is inspired from~\cite{garbin2021fastnerf} and allowed us to learn using a smaller network.

We perform Sobolev training~\cite{czarnecki2017sobolev} of the network using the following loss function:
\begin{align}
	\mathcal{L}= &||T_{\theta}-T_{data}|| + w_{x}||\partial_x T_{\theta} - (\partial_x T)_{data}|| + w_{y}||\partial_y T_{\theta} - (\partial_y T)_{data}|| + \nonumber\\
	&w_t||\partial_t T_{\theta} - (\partial_t T)_{data}||, \label{eq7}
\end{align}
where $T_{\theta}$ is the learned temperature, and $\partial_x T_{\theta}$, $\partial_y T_{\theta}$, and $\partial_t T_{\theta}$ are its derivatives with respect to $x$, $y$, and $t$, respectively, whereas $T_{data}$, $(\partial_x T)_{data}$, $(\partial_y T)_{data}$, $(\partial_t T)_{data}$ are provided in the dataset. For the loss weights, we set $w_{x}=w_{y}=w_{t}=0$ for the first 50 epochs of training to focus on learning only the temperature field. After that, we change them to $w_{x}=5$, $w_{y}=1$ and $w_t=5$.

To assess the accuracy of MeltpoolINR, we also train a CNN-based model. In the CNN, the temperature and its $x$, $y$, and $t$ derivatives are the four independent network outputs, while the loss function remains the same as in~\eqref{eq7}. By treating the temperature and its derivatives as independent outputs, we allow the CNN to potentially overfit the data. The CNN, resembling the network in~\cite{ogoke2023convolutional}, comprises an MLP with 2 hidden layers and 1024 neurons followed by 5 sets of layers, each consisting of upsampling, convolution, and activation layers. We utilize Leaky ReLU as the activation function. See~\ref{net_arch} for the details of the architecture. The CNN outputs a 4-channel field of $128 \times 64$ pixels, the four channels being the temperature and its $x$, $y$, and $t$ derivatives. For this network, we set $w_{x}=w_{y}=w_t=0.1$ in the loss function in~\eqref{eq7}.

In training, we employ the ADAM optimizer with a learning rate of $5\times 10^{-4}$. We use MSE loss, as expressed below:
\begin{equation}
	||u|| = \frac{1}{n_{P3} n_{P2} n_{P1} n_{t} n_{y} n_{x}}\sum_{n=0}^{n_{P3}} \sum_{m=1}^{n_{P2}} \sum_{l=1}^{n_{P1}} \sum_{k=1}^{n_{t}} \sum_{j=1}^{n_{y}} \sum_{i=1}^{n_{x}} u(x^i,y^j,x_l^k,P_1^l,P_2^m,P_3^n)^2,
\end{equation}
where $n_{x},~n_{y},~n_{t},~n_{P1},~n_{P2}$, and $n_{P3}$ are the numbers of grid points in the $x$, $y$, $x_l$, $P_1$, $P_2$, and $P_3$ directions, respectively. Also, for the CNN, $x_l$ is replaced with $t$ above. We randomly separate the data into training and testing cases, and use 90\% of the data for training and the remaining 10\% for testing.

\subsection{Level set method}
\label{lsm}
We aim to learn the geometry of the melt pool, its evolution over time, and its dependence on the process parameters. To accomplish this, we utilize an implicit representation of the melt pool boundary, expressed as the zero level-set of a level-set function $\phi(x,y,t,\mathbf{P})$, where $x$, $y$ and $t$ are the spatial and temporal coordinates, and $\mathbf{P}$ denotes the vector of the process parameters. The evolution of $\phi$ is governed by the following equation~\cite{sethian1999level}:
\begin{equation}
	\frac{\partial \phi}{\partial t}+F|\nabla \phi|=0, \label{eq1}
\end{equation}
where $F=F(x,y,t,\mathbf{P})$ is the speed function governing the evolution of the level set, $\nabla$ denotes the spatial gradient, and $|\cdot|$ is the Euclidean norm. When $\phi$ is known and $|\nabla \phi|\ne 0$, \eqref{eq1} can be used to compute the speed function through:
\begin{equation}
	F = -\frac{1}{|\nabla \phi|}\frac{\partial \phi}{\partial t}. \label{eq2}
\end{equation}

Since the isotherms at the solidus ($T_S$) and liquidus ($T_L$) temperatures define the melt pool boundary and the location of initiation of the solidification process, respectively, we define the level-set function using the temperature field as follows:
\begin{equation}
	\phi=\frac{T}{T_0}-1, \label{eq3}
\end{equation}
where one can set $T_0=T_S$ to track the melt pool and $T_0=T_L$ to track the initiation of the solidification process. The speed function $F$ in the part of the isotherm at $T_L$, where the melt is solidifying, equals the solidification rate. For simplicity, we set $T_0=(T_S+T_L)/2$ in this work.

The printing process can be optimized by controlling the melt pool geometry. The gradient-based optimization of the process parameters then requires the computation of the rate of change of the melt pool geometry and of its relevant features (i.e. width, length, aspect ratio) at a given time with respect to the process parameters. This computation is also facilitated by the level-set approach described by the following equation:
\begin{equation}
	\frac{\partial \phi}{\partial P_i}+F^P_i|\nabla \phi|=0, \label{eq4}
\end{equation}
where $P_i$ is the $i_{th}$ process parameter and $F^P_i\Delta P_i$ is the change in the melt pool boundary at the given time if one chooses $P_i+\Delta P_i$ as the process parameter instead of $P_i$ (where $\Delta P_i$ is small). Consequently, the rate of change of the length ($l$), width ($w$), and aspect ratio ($AR=w/l$) of the melt pool with a change in the process parameter $P_i$ can be estimated via:
\begin{align}
	&\frac{\partial l}{\partial P_i} = F^P_i(x_{\max}, y, t, \mathbf{P})+F^P_i(x_{\min}, y, t, \mathbf{P}), \label{eq5a} \\
	&\frac{\partial w}{\partial P_i} = F^P_i(x, y_{\max}, t, \mathbf{P})+F^P_i(x, y_{\min}, t, \mathbf{P}), \label{eq5b} \\
	&\frac{\partial (AR)}{\partial P_i} = \frac{1}{l}\left[\frac{\partial w}{\partial P_i} - AR \frac{\partial l}{\partial P_i}\right], \label{eq6}
\end{align}
where we are assuming that the printing direction is parallel to the $x$ axis (Figure \ref{fig:sim_model}), $x_{\max}$ and $x_{\min}$ are the $x$-coordinates of the extremes of the melt pool in the $x$ direction, and $y_{\max}$ and $y_{\min}$ are the $y$-coordinates of the extremes of the melt pool in the $y$ direction.


\section{Results}
\label{result}
This section presents the assessment of our proposed model and demonstrates the inference of the change in the melt pool geometry with the process parameters.

\subsection{Assessment of the model}
We evaluate the goodness of fit for field quantities using the coefficient of determination ($R^2$) values. To assess the accuracy of the prediction of the melt pool geometry, we utilize the Chamfer distance metric $-$ see~\ref{cd}.
\begin{figure}[!ht]
	\centering
	\includegraphics[width=0.79\textwidth,valign=t]{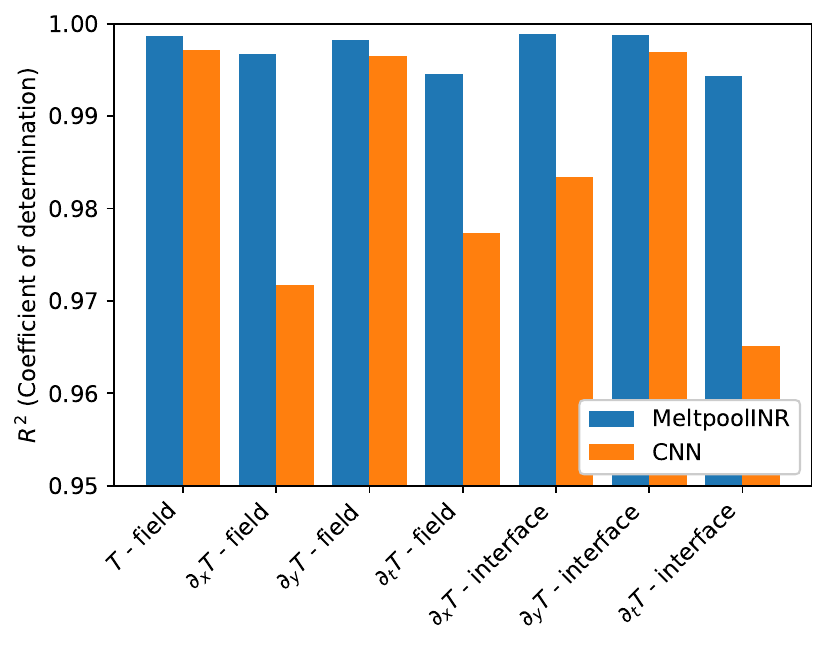} 
	\hfill
	\includegraphics[width=0.195\textwidth,valign=t]{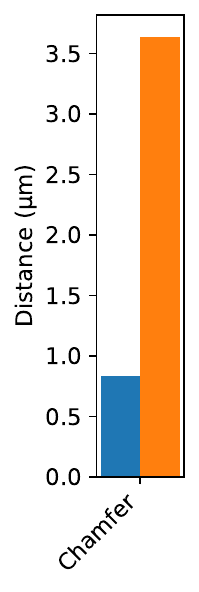}
	\caption{Goodness of fit of the temperature field and its derivatives for the entire domain and at the melt pool boundary i.e. interface (left); Chamfer distance between the melt pool boundary in the data and the prediction (right).
	}
	\label{fig:overall_metrics}
\end{figure}

Figure~\ref{fig:overall_metrics} shows the evaluation metrics computed on the test data set. Both models learn the temperature field accurately, but MeltpoolINR performs better than CNN in learning the temperature gradient and cooling rate, even though the CNN learns the temperature field and its spatial and temporal gradients as independent fields. Additionally, the gradients at the melt pool boundary are assessed. Here too, the performance of MeltpoolINR is better than the CNN. Furthermore, a comparison of the Chamfer distances for the two models shows that MeltpoolINR outperforms CNN. Even though the $R^2$ values for the temperature field predicted by MeltpoolINR and the CNN are comparable, the CNN performs comparatively poorly in learning the melt pool geometry.

\begin{figure}[!ht]
	\centering
	\begin{tabular}{c@{\hspace{0.3cm}}c}
		\raisebox{1.8cm}{\rotatebox[origin=t]{90}{\normalsize $T$}} & \includegraphics[width=0.8\textwidth]{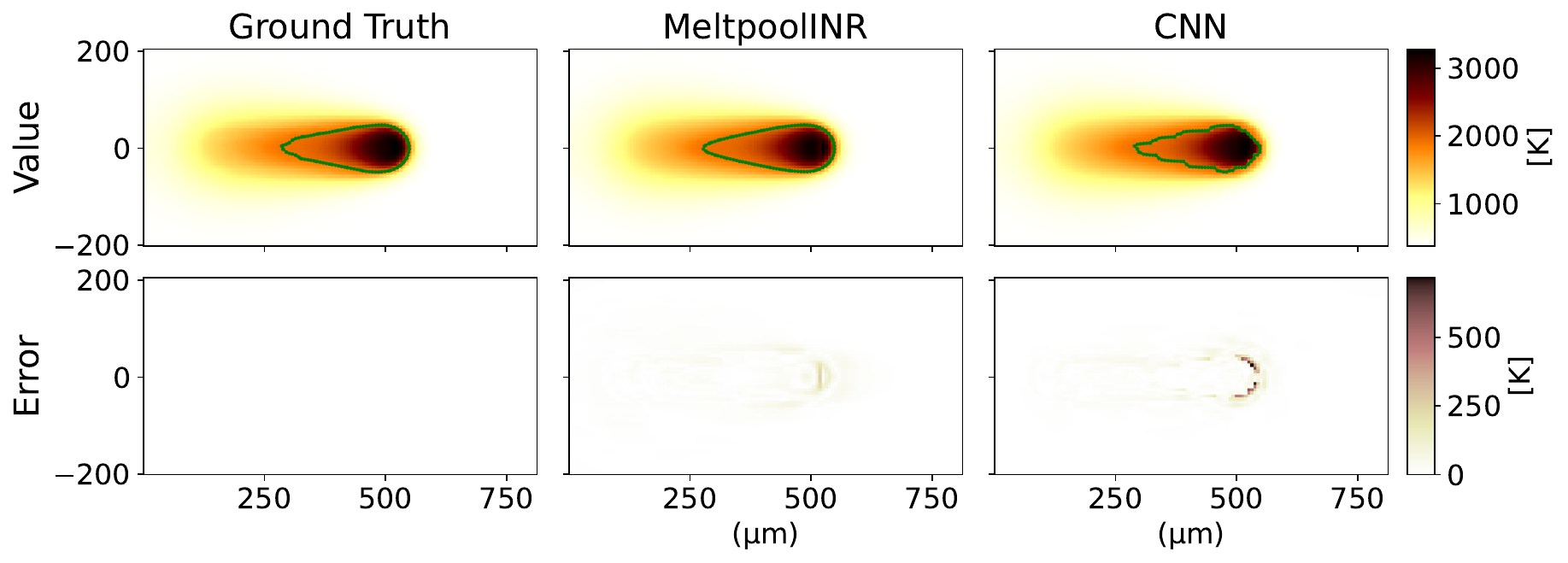} \\
		\raisebox{1.8cm}{\rotatebox[origin=t]{90}{\normalsize  $\partial_x T$ }} & \includegraphics[width=0.8\textwidth]{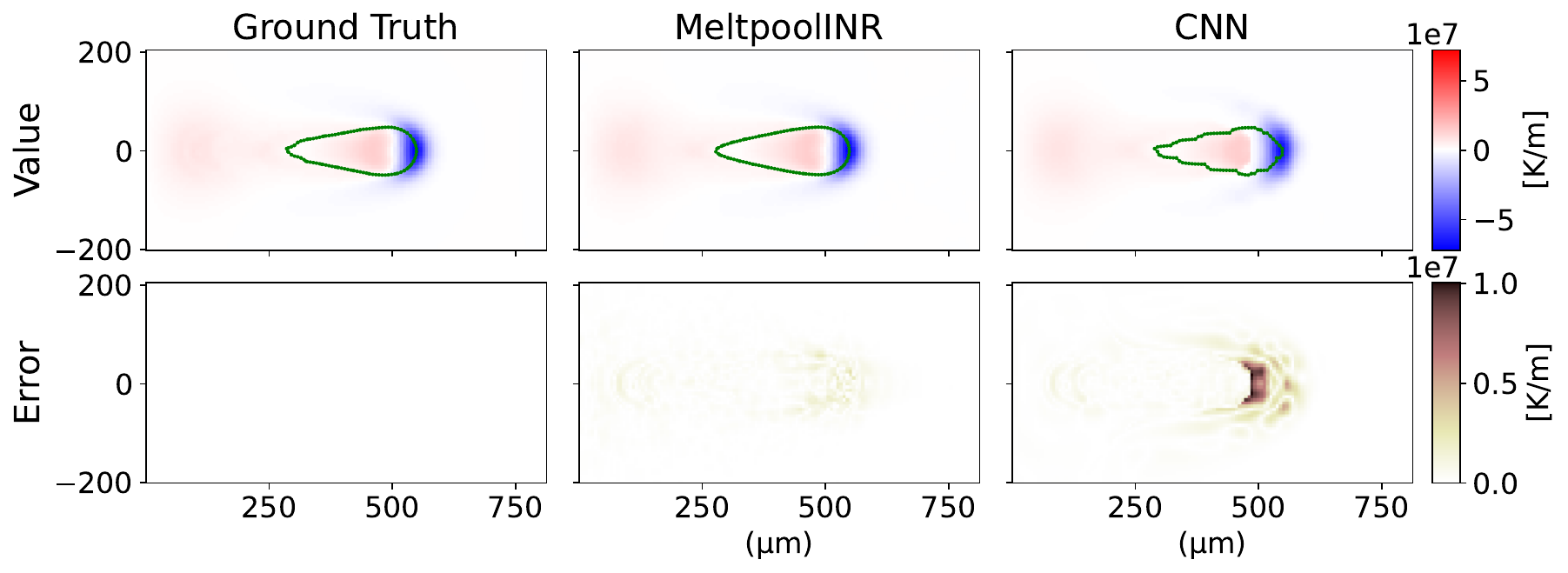} \\
		\raisebox{1.8cm}{\rotatebox[origin=t]{90}{\normalsize  $\partial_y T$ }} & \includegraphics[width=0.8\textwidth]{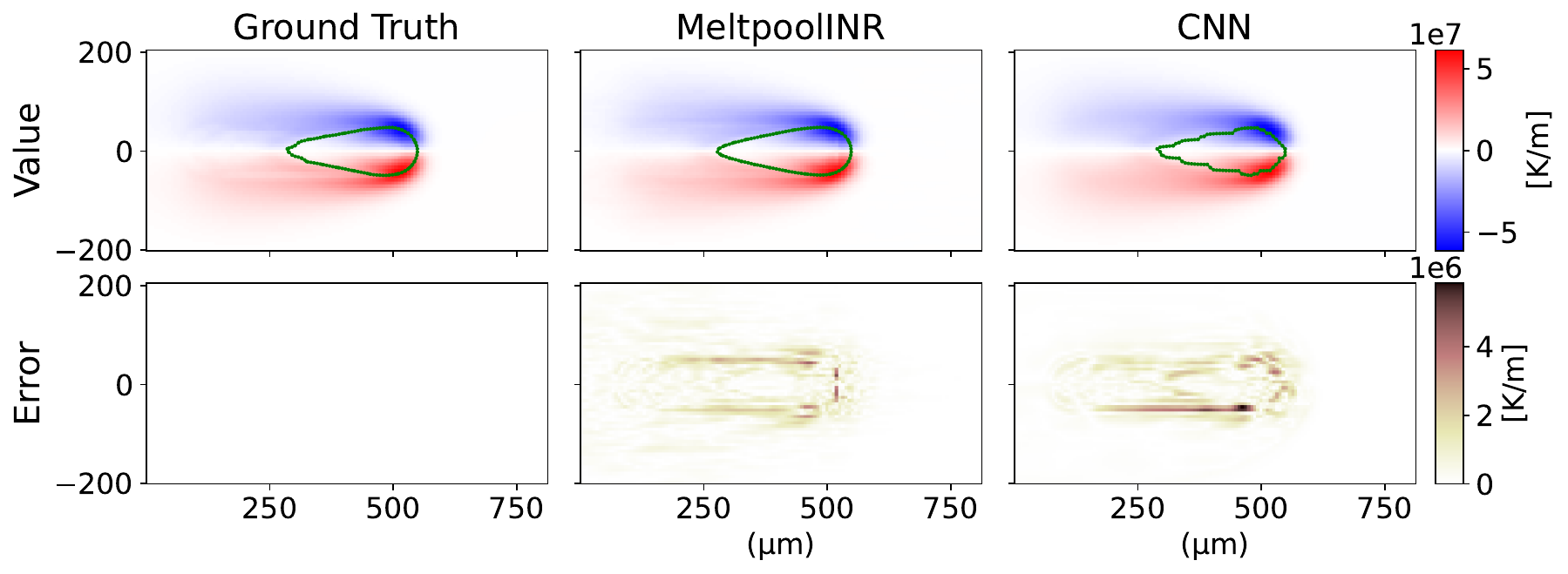} \\
		\raisebox{1.8cm}{\rotatebox[origin=t]{90}{\normalsize  $\partial_t T$ }} & \includegraphics[width=0.8\textwidth]{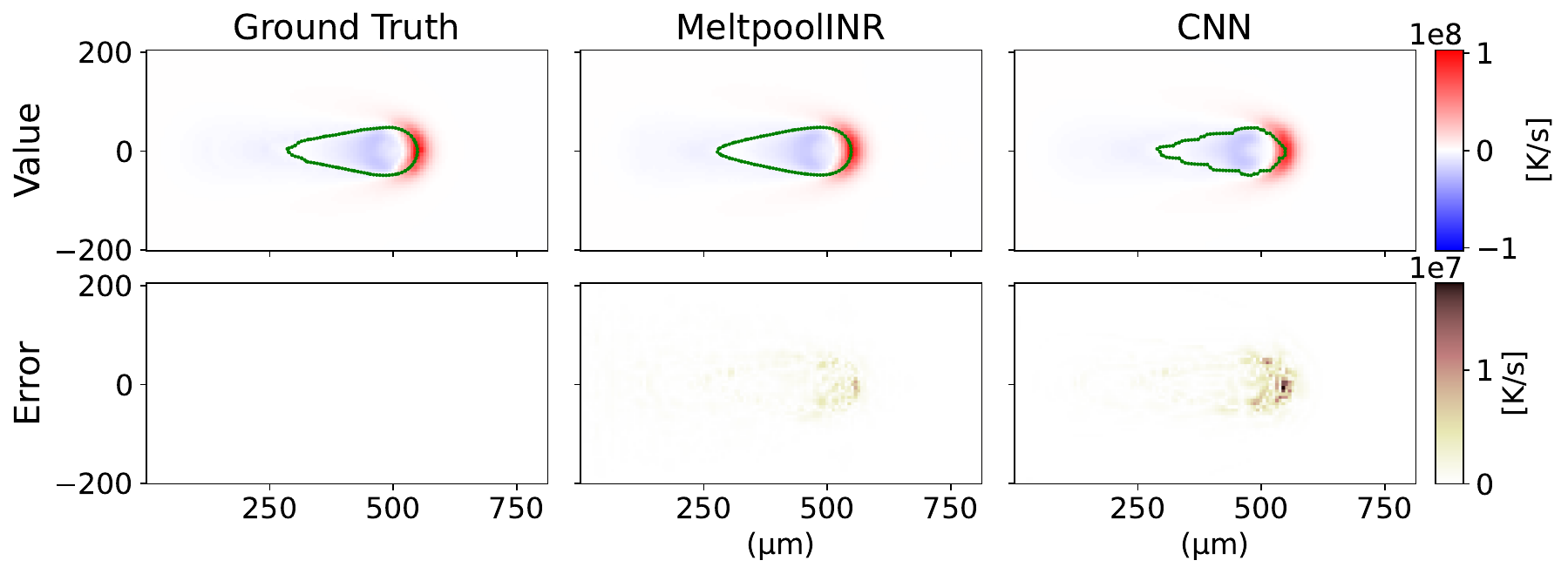} \\
	\end{tabular}
	\caption{Temperature field and its derivatives at time 202$\mu$s for a substrate preheating temperature of 380 K, laser power 160 W, and scanning speed 1.25 m/s. The difference between the ground truth and the outputs of the networks is also shown as an error.}
	\label{fig:output_visualisation}
\end{figure}
Next, we investigate the spatial distribution of errors within the domain. Figure~\ref{fig:output_visualisation} compares the network outputs for a test case with the ground truth obtained from the SPH simulation. The error, defined as the difference between the network output and the ground truth, is also shown. We observe that while  MeltpoolINR learns the melt pool boundary correctly, the CNN exhibits a pronounced step-like error profile. The fields obtained from the CNN have a comparatively higher error, as expected based on the goodness of fit shown in~\fref{fig:overall_metrics}. Furthermore, the errors are localized. As expected, the parts of the domain with steeper slopes of the temperature field are challenging to learn and errors are localized in those regions.

\begin{figure}[ht!]
	\centering
	\includegraphics[width=\textwidth]{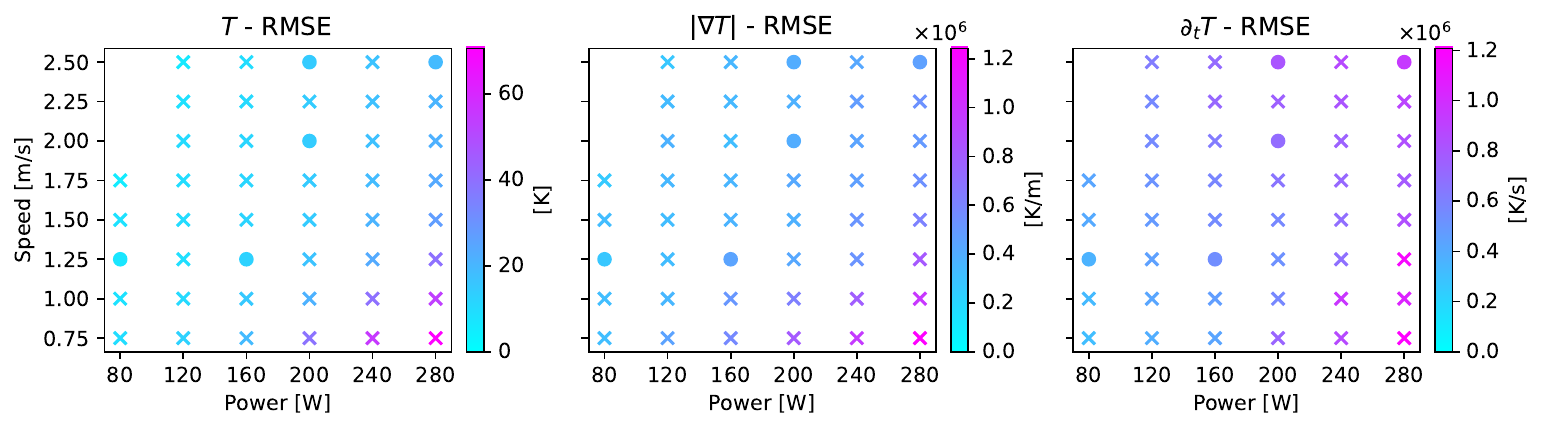}
	\caption{MeltpoolINR: RMSE in the temperature field, its gradient, and the cooling rate at different laser power and scanning speeds for the initial substrate temperature of 380 K. Crosses denote training data, and circles denote test data.
	}
	\label{fig:parameter_exploration}
\end{figure}
To investigate the error variation with the choice of the process parameters, we focus our attention on MeltpoolINR. \fref{fig:parameter_exploration} shows the root mean squared error (RMSE) in the temperature field, and in its spatial and temporal gradients (see~\ref{plots} for the errors for the complete set of process parameters). We observe that the errors in the test cases (shown by filled circles) are similar in magnitude to those of the surrounding training cases. This is an indicator of the generalizability of our model. The error increases for scenarios characterized by high power and low speed. This can be attributed to a qualitative change in the temperature profile indicated by the change in the melt pool shape. In contrast to other cases, the melt pool at high power and low speed is narrower at the front and wider at the back (see~\fref{fig:field_process_param}).

\begin{figure}[ht!]
	\centering
	\includegraphics[width=\textwidth]{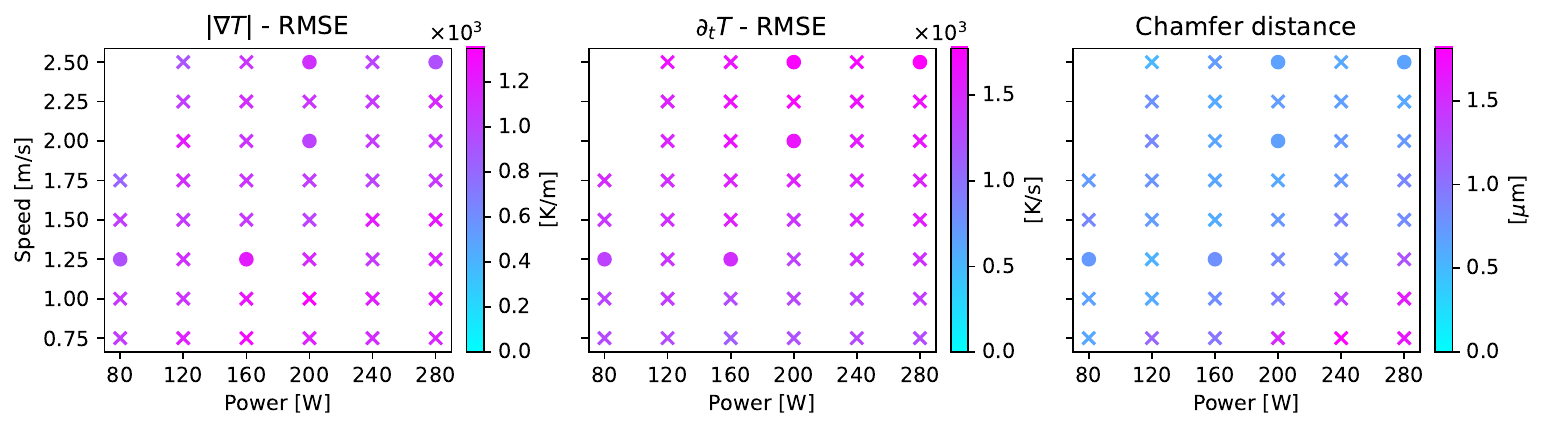}
	\caption{MeltpoolINR: RMSE in the temperature gradient and in the cooling rate at the melt pool boundary together with the error in the melt pool boundary at different laser power and scanning speeds for the initial substrate temperature of 380 K. Crosses denote training data, and circles denote test data.}
	\label{fig:interface_error_persample}
\end{figure}
We also assess the model performance in learning the melt pool boundary, as well as the temperature gradient and the cooling rate at the boundary, as shown in~\fref{fig:interface_error_persample}. A comparison of the scale of the error bars between \fref{fig:parameter_exploration} and \fref{fig:interface_error_persample} reveals that the temperature gradient and the cooling rate at the boundaries are learned with significantly better accuracy. This observation is consistent with the spatial distribution of errors shown in~\fref{fig:output_visualisation}, which reveals that the errors are localized inside the melt pool for MeltpoolINR. Moreover, these errors do not exhibit a strong dependence on the laser power and scanning speed, unlike in~\fref{fig:parameter_exploration}. However, the variation of the errors in the melt pool geometry with laser power and speed exhibits a similar pattern as observed in~\fref{fig:parameter_exploration}.

\subsection{Inference of the melt pool features}
\begin{figure}[!h]
	\centering
	\includegraphics[width=0.49\textwidth]{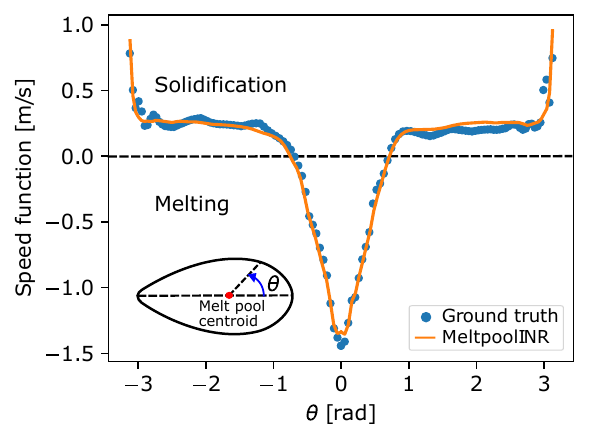}
	\includegraphics[width=0.49\textwidth]{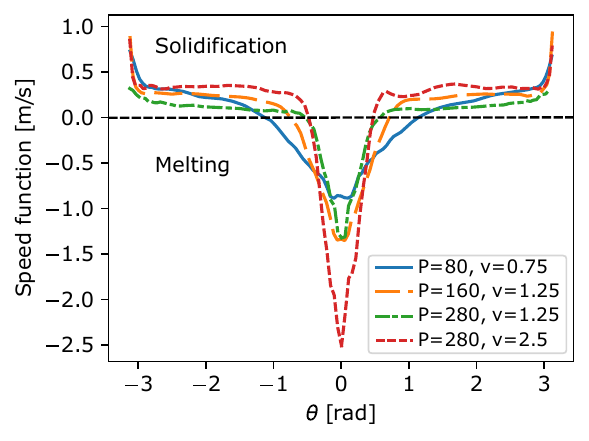}
	\caption{MeltpoolINR: speed function at the melt pool boundary for a laser power of 160 W, a scanning speed of 1.25 m/s, and an initial substrate temperature of 380 K at time 202 $\mu$s (left); Speed function at the melt pool boundary for four different combinations of process parameters (right).
	}
	\label{fig:speed_estimate}
\end{figure}
In this section, we demonstrate the capability of the model to predict the change in the melt pool geometry by predicting the solidification rate and the change in the melt pool length and aspect ratio when the process parameters are changed. First, we compute the speed function ($F$) at the melt pool boundary using \eqref{eq2}, as shown in \fref{fig:speed_estimate}. The melt pool boundary is parameterized by angle $\theta$, as shown in the inset in \fref{fig:speed_estimate}, to show the variation of $F$. $F$ obtained from the simulation data and MeltpoolINR are in close agreement. Note that the noise in the simulation data arises due to the interpolation from the grid data required to determine the melt pool boundary and to estimate the temperature gradient and cooling rate at the boundary. In contrast, MeltpoolINR does not require such interpolation. $F$ exhibits sharper variation at the front (near $\theta=0$) and back (near $\theta=\pm\pi$) of the melt pool boundary, where the curvature of the boundary is high. Away from these points, it changes slowly. Additionally, $F>0$ indicates the solidification rate. The variation of $F$ with a change in the chosen process parameters is also illustrated in \fref{fig:speed_estimate}. At the smallest laser power and scanning speed, $F$ changes gradually from the front of the melt pool to the back. With increasing power and speed, the melt pool stretches and $F$ plateaus in the two sections of the boundary. Near the front and the back of the melt pool, again $F$ exhibits sharp changes. These variations in the solidification rate, which our model can infer, are crucial for understanding the microstructure evolution in LPBF.

Additionally, we estimate the melt pool length and aspect ratio and investigate their rate of change with respect to laser power, scanning speed, and initial substrate temperature at a given time.
\begin{figure}[!ht]
	\centering
	\includegraphics[width=0.49\textwidth]{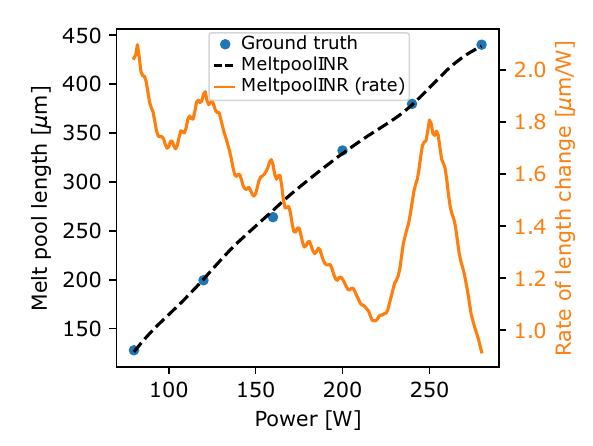}
	\includegraphics[width=0.49\textwidth]{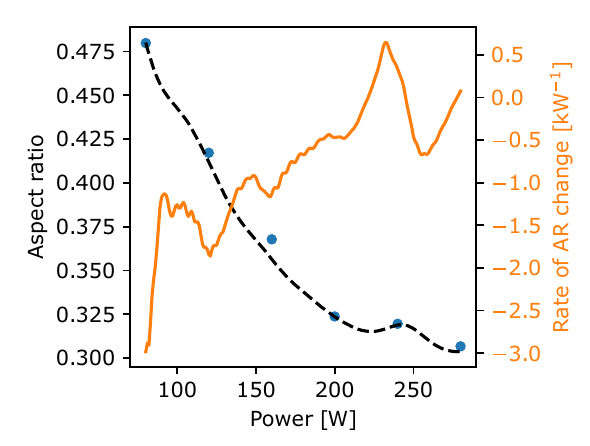}
	\caption{MeltpoolINR: variation of the length and aspect ratio of the melt pool and their rate of change with respect to laser power for a scanning speed of 1.25 m/s and an initial substrate temperature of 380 K at time 202 $\mu$s.}
	\label{fig:MP_features_Pvar}
\end{figure}
\fref{fig:MP_features_Pvar} illustrates their variation with the laser power at a given time, scanning speed, and substrate initial temperature. The values obtained from MeltpoolINR closely align with the ground truth. Increasing power leads to a significant increase in the melt pool length and a considerable decrease in the aspect ratio for most of the power range. The rate of increase of the length lies in a narrow range of values. The rate of change of the aspect ratio for higher values of power indicates that the aspect ratio does not change considerably with increasing power even though the melt pool length increases. This can be attributed to the qualitative change in the melt pool geometry between 240 W and 280 W at a scanning speed of 1.25 m/s observed in~\fref{fig:field_process_param}.

\begin{figure}[!ht]
	\centering
	\includegraphics[width=0.49\textwidth]{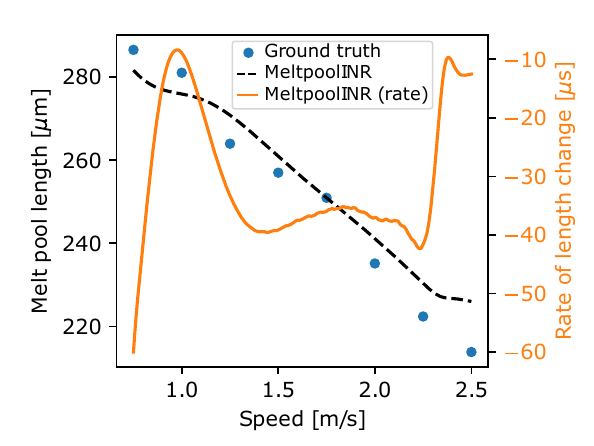}
	\includegraphics[width=0.49\textwidth]{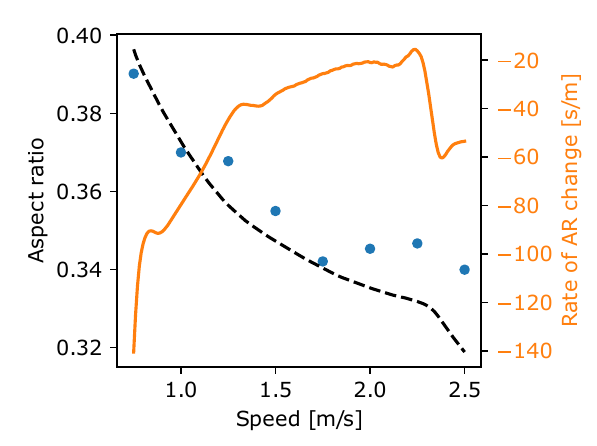}
	\caption{MeltpoolINR: variation of the length and aspect ratio of the melt pool and their rate of change with respect to scanning speed for a laser power of 160 W and an initial substrate temperature of 380 K at time 202 $\mu$s.}
	\label{fig:MP_features_speedvar}
\end{figure}
\fref{fig:MP_features_speedvar} shows the variation of the melt pool length and aspect ratio with respect to scanning speed at a given time, laser power, and substrate initial temperature. The length and aspect ratio decrease with increasing scanning speed. Compared to the case of laser power variation, the change in the length and aspect ratio over the complete range of scanning speed is considerably smaller. However, the rate of the length and aspect ratio change varies over a large range of values. Note that the maximum error in the length and aspect ratio prediction is $\approx 5\%$.

\begin{figure}[!ht]
	\centering
	\includegraphics[width=0.49\textwidth]{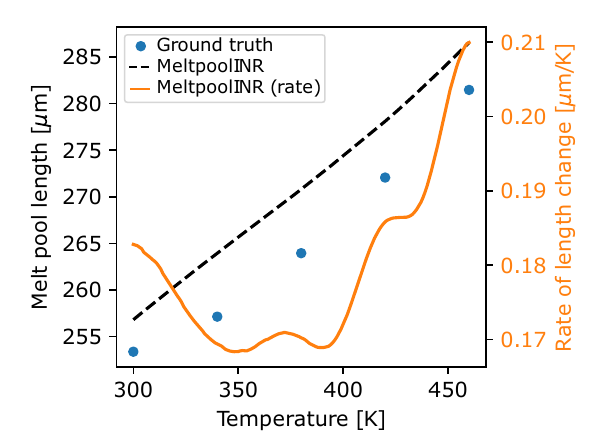}
	\includegraphics[width=0.49\textwidth]{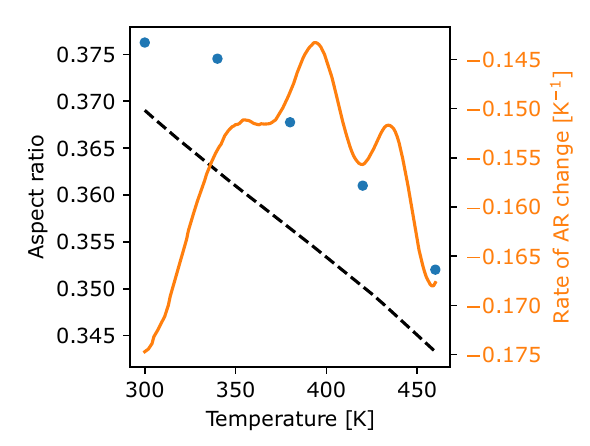}
	\caption{MeltpoolINR: variation of the length and aspect ratio of the melt pool and their rate of change with respect to initial substrate temperature for a laser power of 160 W and a scanning speed of 1.25 m/s at time 202 $\mu$s.}
	\label{fig:MP_features_Tprevar}
\end{figure}
\fref{fig:MP_features_Tprevar} shows the variation of the melt pool length and aspect ratio with respect to substrate initial temperature at a given time, laser power, and scanning speed. The length and aspect ratios increase with increasing substrate initial temperature. However, compared to the cases of the laser power and scanning speed variation, the length and aspect ratio are least sensitive to the variation of the substrate's initial temperature. Note that the maximum error in the length and aspect ratio prediction is $<5\%$.

\section{Discussion}
\label{discussion}
Our aim was to develop a model that accurately predicts the temperature field, melt pool boundary, thermal gradient and cooling rate at the melt pool boundary, and the change in them due to a change in the process parameters in LPBF. We show that MeltpoolINR exhibits strong generalization capability and accurately predicts the temperature field, temperature gradient, and cooling rate as confirmed by~\fref{fig:overall_metrics}. The model also accurately predicts the melt pool boundary as shown in~\fref{fig:overall_metrics}. The implicit neural representation of the melt pool as a level set allows the prediction of the melt pool evolution accurately as shown in~\fref{fig:speed_estimate}. This feature also facilitates the investigation of the dependence of the solidification rate on the process parameters. Furthermore, the model predicts the change in the melt pool geometry with a change in the process parameters in close agreement with the simulation results.

We have demonstrated the effectiveness of our approach in learning the top view of the temperature field, its gradient, and cooling rate in a single-track LPBF process. The desired extension to 3D requires the model to be trained on a dataset with 3D field information without any significant alteration to the model. Further research is needed to extend the model to multi-track printing.

\section{Conclusions}
\label{conclusions}
In the LPBF process, the temperature field, its gradient, and cooling rate along with the melt pool morphology have a critical influence on the properties of fabricated parts. Their quick inference is necessary to model microstructure and residual stresses accurately. Additionally, optimizing the process parameters requires the rate of change of the temperature field and melt pool features with respect to the process parameters. We addressed these challenges in this work by developing and training a differentiable model based on MLPs with Fourier feature encoding. The model predicts a top view of the temperature field, its gradient, and the cooling rate. The accuracy of the prediction is demonstrated by comparison with predictions from a state-of-the-art CNN-based model trained on the same data. Furthermore, by implicitly representing the melt pool boundary as a level set, we quantify the solidification rate and predict the rate of change of the melt pool geometry with respect to the process parameters.

This study focused on learning the top view of fields in a single-track LPBF process to demonstrate the effectiveness of our approach. Future research can extend this framework to 3D geometries for multi-track printing scenarios.

\backmatter

\bmhead{Acknowledgements}
The authors acknowledge the Swiss Data Science Center for funding the project under grant no. C21-07.

\begin{appendices}

\section{Transformation of the frame of reference}
\label{frame_change}
$x,~y$ and $t$ are the spatial and temporal coordinates in the fixed reference frame. We define $\xi,~y$ and $t$ as the coordinates in the frame attached to the laser moving in the $x$-direction of the fixed frame such that 
\begin{align}
\xi &=x-x_l, \label{eq:A1}\\
x_l &=x_0+vt, \label{eq:A2}
\end{align}
where $v$ is the scanning speed, and $x_0$ is the position of the laser at $t=0$. The temperature field is given by $T(x,~y,~t)=\Tilde{T}(\xi,~y,~t)$. The derivatives in the two frames are related by the material derivative as follows:
\begin{equation} \label{eq:A3}
\frac{DT}{Dt}=\frac{\partial \Tilde{T}}{\partial t} - v\frac{\partial \Tilde{T}}{\partial \xi}.
\end{equation}
Note that at steady state, $\frac{\partial \Tilde{T}}{\partial t}=0$. Hence $\frac{DT}{Dt}=- v\frac{\partial \Tilde{T}}{\partial \xi}$. Furthermore, since material particles are stationary in the fixed frame, $\frac{DT}{Dt}$ is the time derivative in the fixed frame, i.e. $\frac{DT}{Dt}=\frac{\partial T}{\partial t}$. In our data, $v$ is constant during a printing process. So, on using \eqref{eq:A2}, \eqref{eq:A3} transforms to the following form:
\begin{equation} \label{eq:A4}
\frac{\partial T}{\partial t}=v \left( \frac{\partial \Tilde{T}}{\partial x_l} - \frac{\partial \Tilde{T}}{\partial \xi}\right).
\end{equation}

\section{Architecture of the CNN}
\label{net_arch}
\begin{figure}[!ht]
    \centering
    \includegraphics[width=0.95\textwidth]{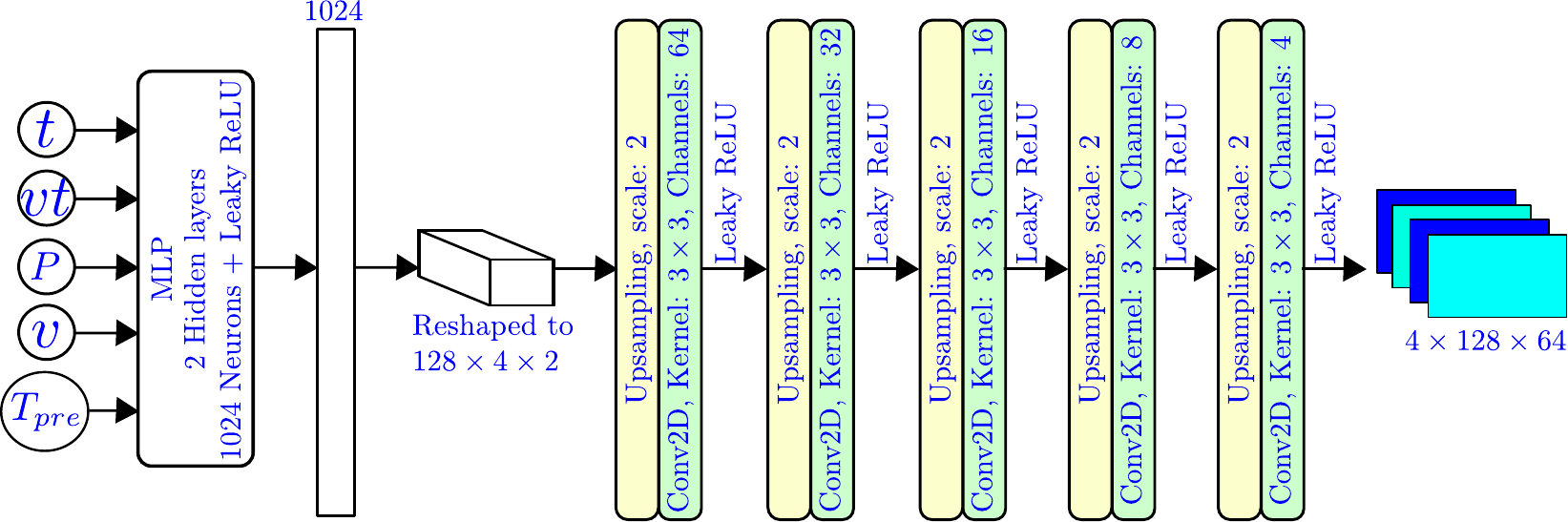}
    \caption{Architecture of the CNN.}
    \label{fig:CNN_architecture}
\end{figure}

\fref{fig:CNN_architecture} shows the architecture of the CNN. Note that upsampling uses the bilinear algorithm. The stride of the convolution is set to be 1. Additionally, a padding of size 1 is added to the four boundaries of the input to a convolution layer by replicating the values at the corresponding boundaries.

\section{Chamfer distance}
\label{cd}
Chamfer distance is a frequently used metric in Computer Vision to compute the distance between two point clouds. In order to calculate it, we resample the true melt pool boundary and the predicted melt pool boundary into two large point clouds $\mathcal{C}_1,~\mathcal{C}_2$ and then compute the Chamfer distance as follows:
\begin{equation} \label{eq:d_chamfer}
d_{\rm{Chamfer}} = \frac{1}{2 N} \left( \sum_{x\in \mathcal{C}_1} \min_{y \in \mathcal{C}_2} \|x -y \|_2 + \sum_{y\in \mathcal{C}_2} \min_{x \in \mathcal{C}_1} \|y - x \|_2 \right)
\end{equation}

\section{RMSE for the complete set of process parameters}
\label{plots}
\begin{figure}[!ht]
    \centering
    \includegraphics[width=0.9\textwidth]{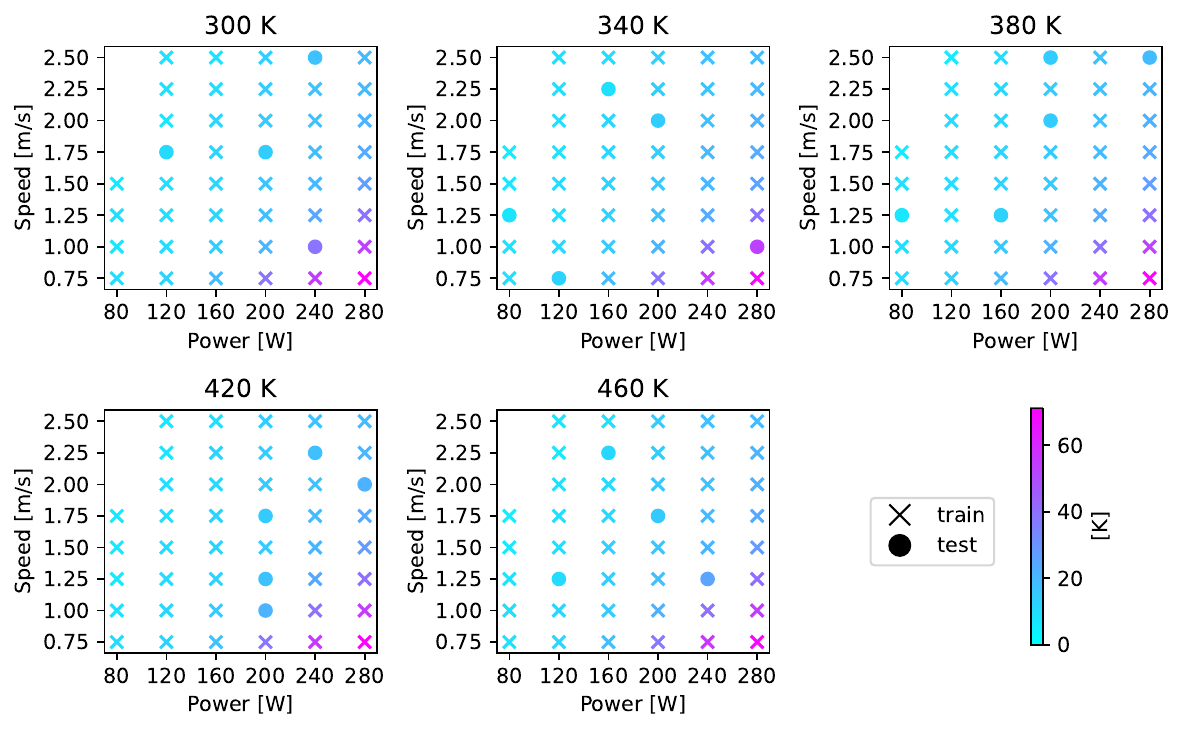}
    \caption{MeltpoolINR: RMSE of the temperature field.}
    \label{fig:parameter_exploration1}
\end{figure}

\begin{figure}[!ht]
    \centering
    \includegraphics[width=0.9\textwidth]{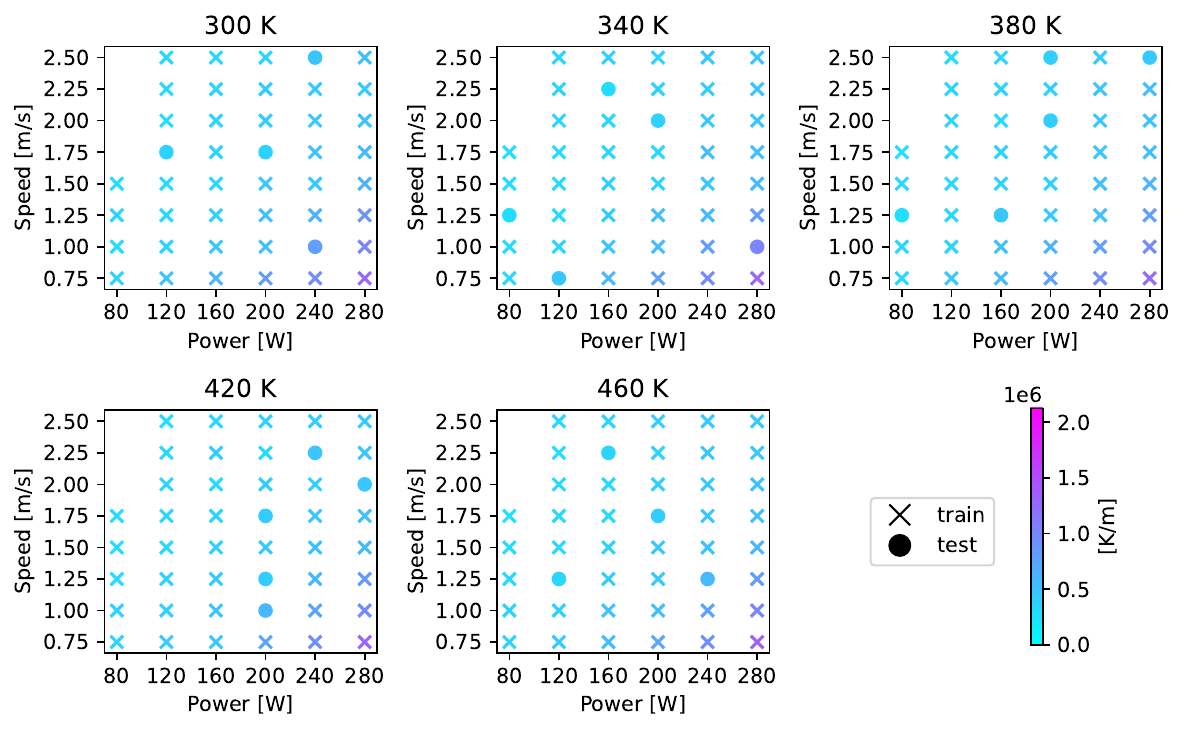}
    \caption{MeltpoolINR: RMSE of the gradient of the temperature field.}
    \label{fig:parameter_exploration2}
\end{figure}

\begin{figure}[!ht]
    \centering
    \includegraphics[width=0.9\textwidth]{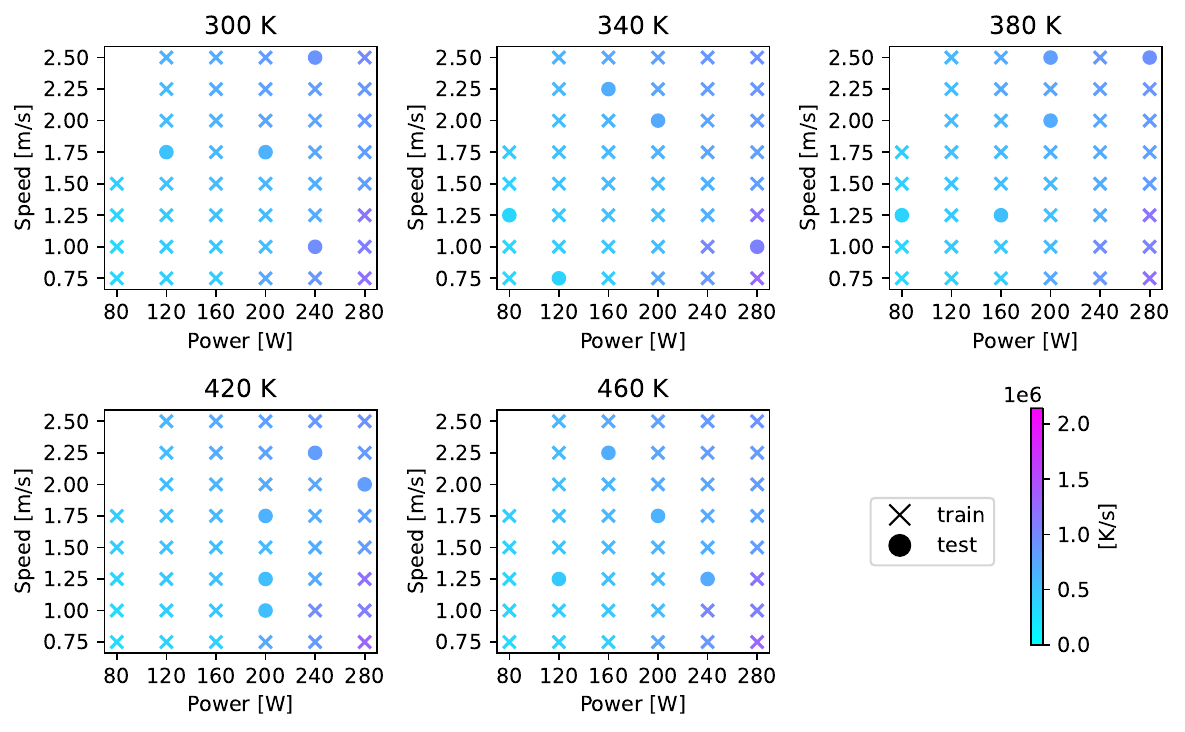}
    \caption{MeltpoolINR: RMSE of the cooling rate.}
    \label{fig:parameter_exploration3}
\end{figure}

\begin{figure}[!ht]
    \centering
    \includegraphics[width=0.9\textwidth]{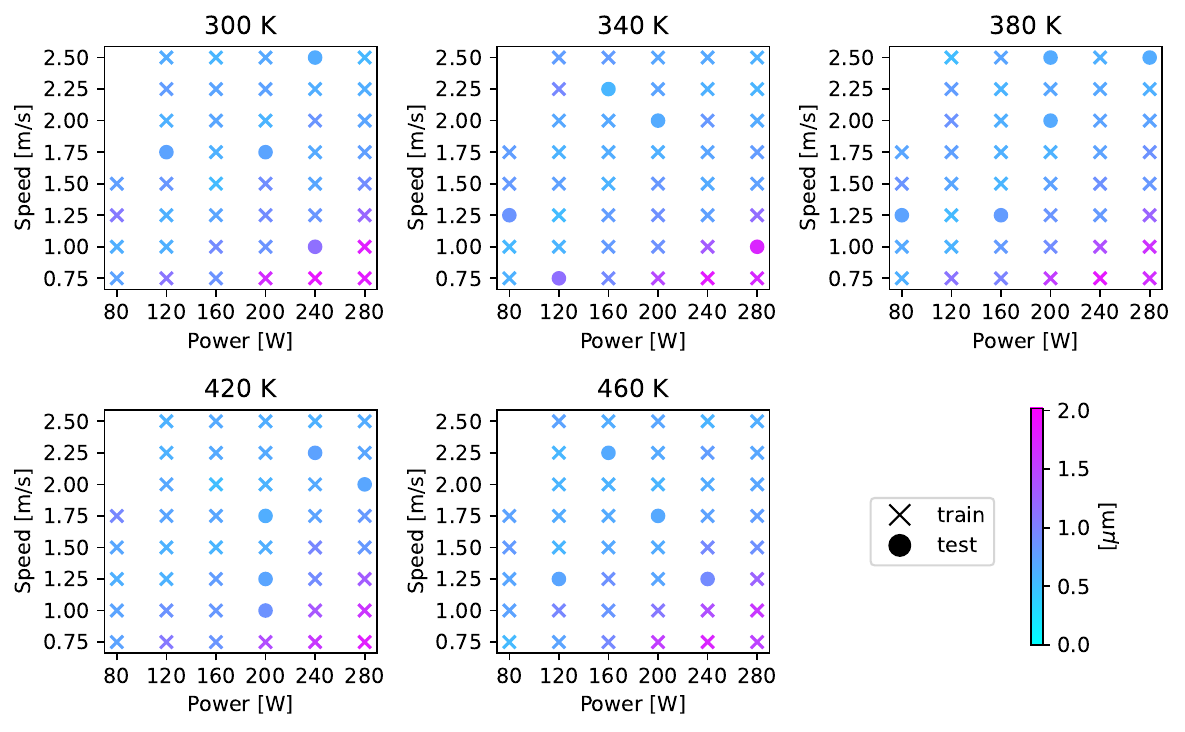}
    \caption{MeltpoolINR: Error in the melt pool boundary measured as the Chamfer distance.}
    \label{fig:cd_error_persample}
\end{figure}

\section{Temperature field in the steady state}
\label{T_v_p}
\begin{figure}[!ht]
    \centering
    \includegraphics[width=0.99\textwidth]{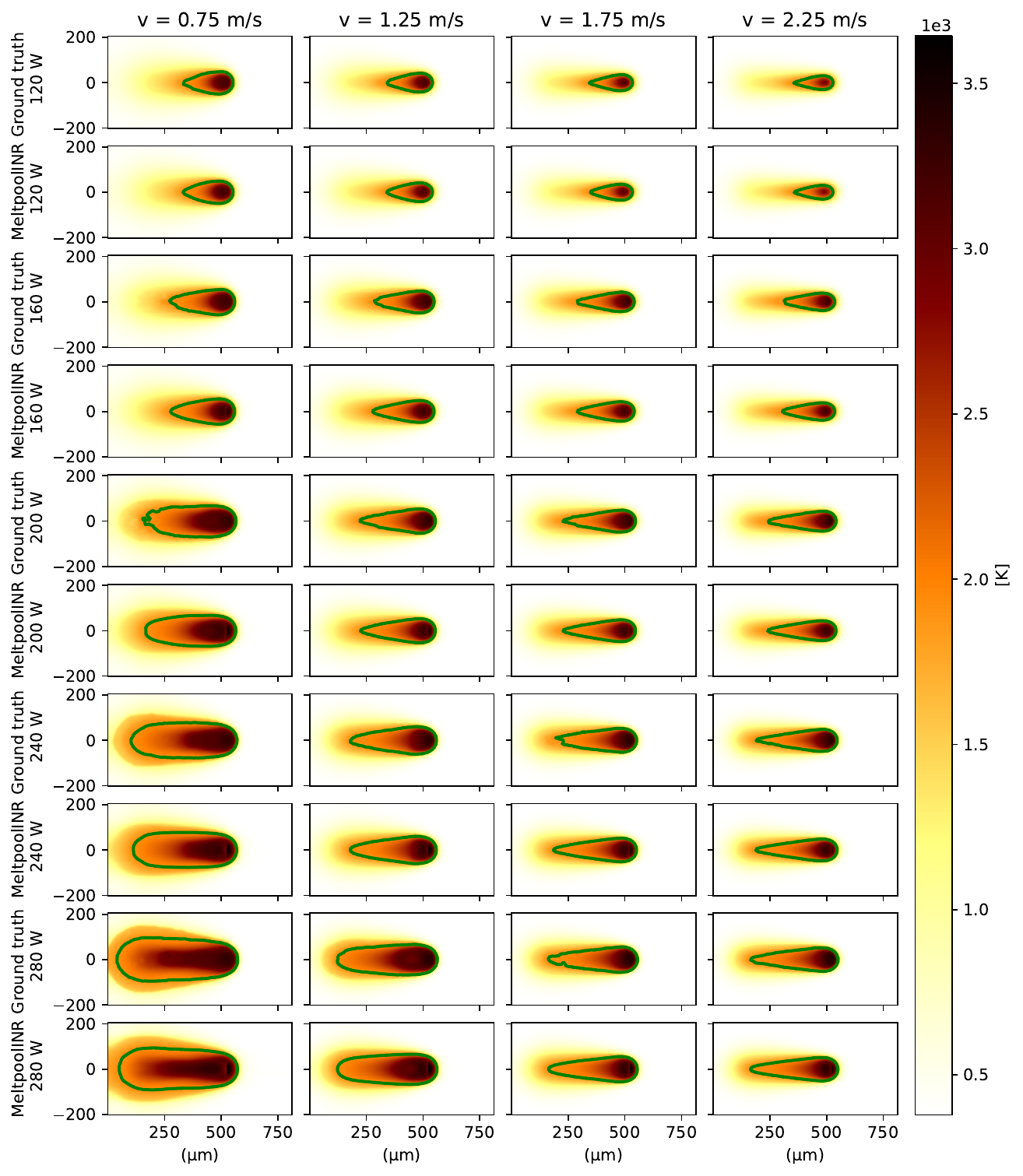} 
    \caption{The temperature field and melt pool boundary at time 202 $\mu$s for different laser power and velocities and an initial substrate temperature of 380 K.}
    \label{fig:field_process_param}
\end{figure}

\end{appendices}

\bibliography{ms}

\end{document}